\documentclass[twocolumn,showpacs,preprintnumbers,amsmath,amssymb]{revtex4-1}

\usepackage{graphicx}
\usepackage{dcolumn}%
\usepackage{bm}%

\begin{document}



\title{Fourier transform spectroscopy, relativistic electronic structure calculation, and coupled-channel deperturbation analysis of the fully mixed $A^1\Sigma^+_u$ and $b^3\Pi_u$ states of Cs$_2$}



\author{A. Znotins, A. Kruzins, M. Tamanis, and R. Ferber}
\email[]{ferber@latnet.lv}
\affiliation{Laser Center, Department of Physics, University of Latvia, 19 Rainis blvd, Riga LV-1586, Latvia}

\author{E. A. Pazyuk, A. V. Stolyarov, and A. Zaitsevskii}
\altaffiliation{Petersburg Nuclear Physics Institute named by B.P. Konstantinov of National Research Center ``Kurchatov Institute'', 188300 Gatchina, Leningrad District, Russia}
\affiliation{Department of Chemistry, Lomonosov Moscow State University, 119991, Moscow, Leninskie gory 1/3, Russia}


\date{\today}

\begin{abstract}
The 4503 rovibronic term values belonging to the mutually perturbed $A^1\Sigma^+_u$ and $b^3\Pi_u$ states of Cs$_2$ were extracted from laser induced fluorescence (LIF) $A\sim b\rightarrow X^1\Sigma^+_g$ Fourier transform spectra with the 0.01 cm$^{-1}$ uncertainty. The experimental term values of the $A^1\Sigma^+_u\sim b^3\Pi_u$ complex covering the rotational levels $J\in [4,395]$ in the excitation
energy range $[9655,13630]$ cm$^{-1}$ were involved into coupled-channel (CC) deperturbation analysis. The deperturbation model takes explicitly into account spin-orbit coupling of the $A^1\Sigma^+_u(A0^+_u)$ and $b^3\Pi^+_{0_u}(b0^+_u)$ states as well as spin-rotational interaction between the $\Omega=0$, $1$ and $2$ components of the $b^3\Pi^+_{\Omega_u}$ state. The \emph{ab initio} relativistic calculations on the low-lying electronic states of Cs$_2$ were accomplished in the framework of Fock space relativistic coupled cluster (FSRCC) approach to provide the interatomic potentials of the interacting $A0^+_u$ and $b0^+_u$ states as well as the relevant $A\sim b$ spin-orbit coupling function. To validate the present CC deperturbation analysis solely obtained by energy-based data, the $A\sim b \to X(v^{\prime\prime}_X)$ LIF intensity distributions were measured and compared with their theoretical counterparts obtained by means of the non-adiabatic vibrational wave functions of the $A\sim b$ complex and the FSRCC $A\sim b \to X$ transition dipole moments calculated by the finite-field method.
\end{abstract}


\maketitle


\section{Introduction}\label{ruvin-intro}

The spectroscopic studies of the heaviest natural alkali dimer, the Cs$_2$ molecule represents a challenging task due to a very dense manifold of vibronic levels, especially concerning the rotational structure. What is more, the first electronically excited states of Cs$_2$ are forming the singlet-triplet $A^1\Sigma^+_u\sim b^3\Pi_u$ complex, which is fully mixed by strong spin-orbit-coupling, thus being by itself a challenge for adequate processing of high resolution spectroscopic information~\cite{Field, Pazyuk:15}.  At the same time there is an obvious necessity to study these states since they have been exploited as the intermediates to transfer the weakly-bound ultracold molecules produced in the $X^1\Sigma^+_g$ state to deeply-bound species in order to produce Cs$_2$ in vibrational and rotational  ultracold "absolute" ground state $X(v = 0, J = 0)$ as it was demonstrated in Refs.\cite{Danzl2010, Danzl2008, Mark2009, Danzl2009, DanzlPhD2010}, in particular, applying the 4-photon stimulated Raman adiabatic passage (STIRAP). One may also mention an efficient accumulation of cold Cs$_2$ molecules in the lowest $X(v = 0)$ level by vibrational cooling~\cite{Viteau2008}. Additional motivation of producing cold Cs$_2$ is its predicted sensitivity for checking the possible variation of the electron/proton mass ratio as discussed in Ref.\cite{DeMille2008}.
The $A\sim b$ window was used for Cs$_2$ to access the triplet state manifold, which makes it possible to implement the perturbation facilitated optical-optical double resonance (OODR)~\cite{Xie2008}; this is particularly important for homonuclear molecules to overcome the restricted choice of available transitions because of the $u-g$ parity selection rule.

The spectroscopic studies of the $A\sim b$ complex of heteronuclear alkali diatomics containing Cs atom have achieved a reasonable progress. Processing Fourier transform (FT) spectroscopy data by the coupled-channel (CC) deperturbation treatment and  applying \textit{ab initio} calculations of spin-orbit coupling functions it was possible to reproduce the large set of experimental data with accuracy better than 0.01 cm$^{-1}$, see \cite{Docenko2010} and \cite{KruzinsRbCs2014} for RbCs, \cite{KruzinsKCs2010} and \cite{KruzinsKCs2013} for KCs, \cite{ZaharovaNaCs} for NaCs and with the accuracy of 0.05 cm$^{-1}$ for LiCs~\cite{Kowalzcyk2015}. In particular, the accuracy and abundance of information on $A\sim b$ system of RbCs allowed us to assign  laser induced fluorescence (LIF) transitions from higher-excited electronic states to the $A\sim b$  complex in case when the respective LIF to the ground $X$-state was too weak to be observed, see Ref.~\cite{Alps2017}. The generated comprehensive set of RbCs $A\sim b$ term values was helpful to predict the $A\sim b\leftarrow X$ transition frequencies to be used for two-step laser excitation of the $(4)^1\Pi$ state~\cite{Klincare2018}. For the closest homonuclear analogue of Cs$_2$, the rubidium dimer~\cite{Salami2009, Drozdova2013}, the 4500 rovibronic term values covering 93\%\ of the $A$ state well depth were reproduced with a standard deviation of 0.005 cm$^{-1}$ matching the experimental uncertainties below 0.01 cm$^{-1}$.

Regarding Cs$_2$, the first detailed experiment based study of the $A\sim b$ complex was performed in Ref.~\cite{Bai2011}, in which the information from several sources obtained by different spectroscopic methods that included but are not restricted to the data from Refs~\cite{Verges1987, Amiot2002, Xie2008, SuplPRA83}, was processed by a deperturbative CC approach. The data however have been sparse and a considerable range of energies and rotational levels $J$ of the $A\sim b$ complex was not covered, therefore, as concluded in Ref.~\cite{Bai2011}, there are issues that warrant further study.

\begin{figure}

\includegraphics[scale=1]{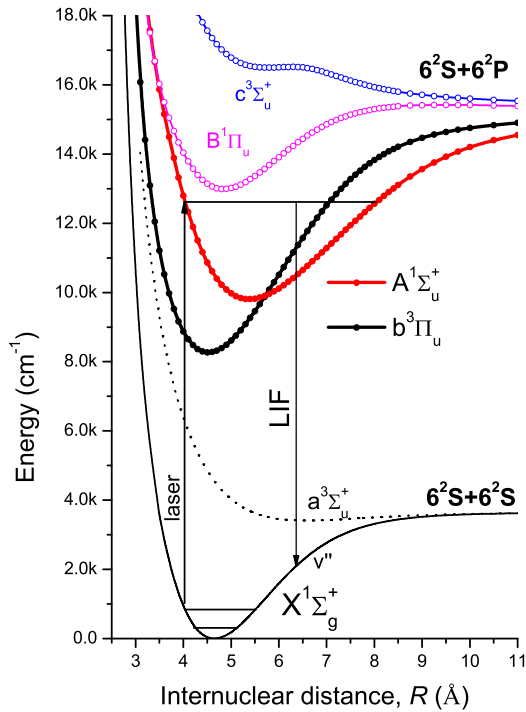}
\caption{(Color online) The scheme of the scalar-relativistic potential energy curves~\cite{Pazyuk2015} of Cs$_2$ converging to the lowest two non-relativistic dissociation limits. Only $u$-symmetry excited states are presented.}\label{Fig_PEC}
\end{figure}

There is quite a number of papers reporting on the \textit{ab initio} electronic structure calculations on adiabatic potential energy curves (PECs), see Fig.~\ref{Fig_PEC}, and transition dipole moments of Cs$_2$, which have been performed in the framework of pure Hund's (\textbf{a}) and (\textbf{c}) coupling cases~\cite{Bai2011, Krauss:90, Foucrault:92,  Allouche2012, Pazyuk2016, Zaitsevskii:17}. The individual spin-orbit (SO) coupling matrix elements between the scalar-relativistic electronic states were calculated in Ref.~\cite{Pazyuk2015}.

The strategy of the present study was as follows. As distinct to Ref.~\cite{Bai2011}, we intended to base mainly on the rotationally resolved by FT spectroscopy $A\sim b\rightarrow X$ LIF measurements performed in the Laser Center in Riga (University of Latvia) with the accuracy of 0.01 cm$^{-1}$ or better. By making all possible use of LIF excitation accessible with a Ti-Sapphire laser and diode lasers at our disposal and due to collision-induced population of rotational levels it was possible to obtain the systematically spanned over $J$ data field and to pass from rather scarce amount of fragmentary data to much more uniform sufficiently dense coverage of term values extended towards higher energies of about 13 500 cm$^{-1}$. To include the term values below the singlet $A$-state minimum we corroborated the present data by the only available $b^{3}\Pi_{0u}^+$ state term values from two-photon excitation experiments used in Ref.~\cite{Bai2011}, though of poor (monochromator) accuracy. To propagate properly the experimental data to the lowest $J$ values ($J$ = 1 and 3) the 19 accurate term values of the $A\sim b$ complex used in Ref.~\cite{DanzlPhD2010} (University of Innsbruck) for the STIRAP assembling of ultracold Cs$_2$ molecules were involved in the present fit as well. This entire set of data have been treated in the framework of the rigorous deperturbation analysis based on the four coupled-channel $A^1\Sigma^+_u\sim b^3\Pi^+_{u(0,1,2)}$ model. To probe the CC deperturbated parameters obtained only by energy-based data we compared the experimental $A\sim b \to X(v^{\prime\prime}_X)$ LIF relative intensity distributions with their calculated counterparts. For this purpose the \textit{ab initio} transition dipole moments were evaluated in the present study by means of the finite-field (FF) method combined with multi-reference Fock space relativistic coupled cluster (FSRCC) method~\cite{Visscher:01, Zaitsevskii:17, Zaitsevskii:18}. To additionally validate the FSRCC potentials we have estimated \emph{ab initio} $\Omega$-doubling effect in the low-lying vibrational levels of the $b^3\Pi^{\pm}_{0_u}$ state previously measured in Ref.~\cite{Xie2008}.

\section{Experiment}\label{maris-experiment}

\subsection{Experimental setup}
The rovibronic term values of the mixed $A^1\Sigma^+_u$ and $b^3\Pi_u$ states in Cs$_2$ were obtained from Fourier transform spectra of laser induced fluorescence $A\sim b\rightarrow X$ recorded with the Bruker IFS -125HR spectrometer. Cs$_2$ molecules were produced in a linear heat-pipe at about 300$^{\rm o}$C. For detection of near infra-red LIF the InGaAs diode was used. About a half of the data was collected from previous experiments on the $A\sim b$ complex of the Cs containing heteronuclear dimers KCs and RbCs \cite{Docenko2010, KruzinsRbCs2014, KruzinsKCs2010, KruzinsKCs2013, Tamanis2010, Docenko2011} since the respective molecular vapor in the heat-pipe contained Cs$_2$ molecules as well. Due to the accidental excitations of the latter, Cs$_2$ $A^1\Sigma^+_u\sim b^3\Pi_u\rightarrow X^1\Sigma^+_g$ transitions were also observed in the recorded spectra. Most of these spectra were excited with home-made external cavity diode lasers with 980, 1020, and 1050 nm laser diodes. To obtain a more systematic data set the heat-pipe containing only cesium metal was manufactured; and a Ti-Sapphire laser (Coherent MBR 110) was used to excite the  Cs$_2$ $A\sim b$ complex. The exploited laser frequencies in this case were within 10 746 -- 11 612 cm$^{-1}$.

\subsection{Spectra analysis}
\begin{figure}
\includegraphics[scale=0.8]{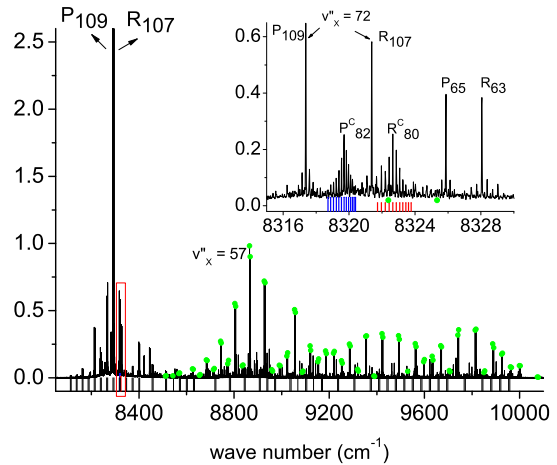}
\caption{(Color online) The  $A\sim b(J^{\prime}; E_{A\sim b}) \rightarrow X(J^{\prime\prime};v^{\prime\prime}_X)$ LIF spectrum of Cs$_2$ recorded at excitation $\nu_{laser}$ = 10 746.02 cm$^{-1}$. Two strongest progressions originate from levels $J^{\prime}$ = 81, $E_{A\sim b}$ = 11 053.533 cm$^{-1}$, see green solid cycles on top of the lines, and $J^{\prime}$ = 108, $E_{A\sim b}$ = 10 990.136 cm$^{-1}$, see bars below the spectrum. An edge filter FEL1000, placed in observation path, cuts off transitions higher than 10 000 cm$^{-1}$. The inset shows zoomed in fragment of the spectrum with collissionally induced fluorescence (CIF) from $b^{3}\Pi_{0u}^+$, see $P^C$, $R^C$ line groups. The strongest $P^C$, $R^C$ lines originate from the level $J^{\prime}$ = 81, $E_{A\sim b}$ = 11 050.848 cm$^{-1}$. Two green solid cycles in the inset show calculated position of $P$, $R$ doublet at $v^{\prime\prime}_X$ = 76 from optically excited level $J^{\prime}$ = 81. The indices denote a ground state rotational level $J^{\prime\prime}=J^{\prime}\pm 1$.}\label{Fig_sp}
\end{figure}

The rotational and vibrational assignment of LIF progressions was based on a comparison of the observed  vibrational-rotational differences with their calculated counterparts obtained  using highly accurate empirical potential available for the ground $X^1\Sigma^+_g$ state~\cite{Coxon:10}. The LIF spectra were rather dense and typically contained more than ten doublet $P$, $R$ progressions. Due to the presence of buffer gas Ar in the heat-pipe the rotational relaxation lines were observed around strong spectral lines, thus a large amount of term values of collisionally populated levels could be obtained. In several spectra the population transfer from an optically excited rovibronic levels to the $b^{3}\Pi_{0u}^+$ state was observed. This is illustrated, see green solid cycles on top of the lines, in Fig.~\ref{Fig_sp}. A strong progression with maximal intensity at $v^{\prime\prime}_X$ = 57 originates from the level with $J^{\prime}$ = 81, $E_{A\sim b}$ = 11 053.533 cm$^{-1}$. This progression ends at $v^{\prime\prime}_X$ = 69 ($\nu_{LIF}$ = 8514 cm$^{-1}$)  because of Franck-Condon overlap becomes small for $A\sim b\to X$ fluorescence for high $v^{\prime\prime}_X$. However collisional population transfer from the directly excited level to $b^{3}\Pi_{0u}^+$ rovibronic levels with odd $J^{\prime}$ ranging from 73 to 93 gives rise to transitions from these levels to higher $v^{\prime\prime}_X$. This is shown in the inset of Fig.\ref{Fig_sp} by the fragment of the spectrum, in which the groups of $P^C$, $R^C$- transitions to $v^{\prime\prime}_X$ = 76 are seen. For the strongest lines of these groups the upper state has the same quantum number $J^{\prime}$ = 81 and is shifted by $-$2.7 cm$^{-1}$ from the directly excited level. The collisionally induced fluorescence (hereafter denoted as CIF) of such type was observed in thirteen spectra yielding valuable term value data for the triplet $b^{3}\Pi_{0u}^+$ state.

\section{The coupled-channel deperturbation analysis}\label{avstol-deperturbation}

\subsection{Rovibronic Hamiltonian and fitting procedure}
The spin-orbit coupling between the $A^{1}\Sigma_{u}^{+}$ and $b^{3}\Pi_{0u}^+$ states in the Cs$_{2}$ dimer is larger than the vibrational intervals of the interacting states~\cite{Pazyuk:15}. Therefore, a rigorous coupled-channel (CC) deperturbation treatment is indispensably required to represent the fully mixed rovibronic levels of the singlet-triplet $A\sim b$ complex of Cs$_{2}$ with the experimental (spectroscopic) accuracy.

The rovibronic Hamiltonian~\cite{Docenko2010} of the present deperturbation model takes into account explicitly the dominant SO interaction between the $A^{1}\Sigma_{u}^{+}$ state and the $b^{3}\Pi_{0u}^+$ component, as well as the spin-rotational coupling between the different $\Omega$-components of the triplet $b^3\Pi^{+}_{u(0,1,2)}$ state:
\begin{eqnarray}\label{Ham}
\langle ^{1}\Sigma^{+}|H|^{1}\Sigma^{+} \rangle & = & U_{A0^+_u}+ B(X+2) \nonumber \\
\langle ^{3}\Pi_{0}^+|H|^{3}\Pi_{0}^+ \rangle & = & U_{b0^+_u} + B(X+2)\nonumber \\
\langle ^{3}\Pi_{1}|H|^{3}\Pi_{1} \rangle & = & U_{b0^+_u} + A_{01} + B(X+2)\nonumber \\
\langle ^{3}\Pi_{2}|H|^{3}\Pi_{2} \rangle & = & U_{b0^+_u} + A_{01} + A_{12} + B(X-2)\nonumber \\
\langle ^{1}\Sigma^{+}|H|^{3}\Pi_{0}^+ \rangle & = & - \xi_{Ab0} \\
\langle ^{3}\Pi_{0}^+|H|^{3}\Pi_{1} \rangle & = & - B\sqrt{2X}\nonumber \\
\langle ^{3}\Pi_{1}|H|^{3}\Pi_{2} \rangle & = & - B\sqrt{2(X-2)}\nonumber\\
\langle ^{1}\Sigma^{+}|H|^{3}\Pi_{1} \rangle & = & \eta_{Ab1} B\sqrt{2X} \nonumber
\end{eqnarray}
where $X \equiv J(J+1)$ and $B \equiv \hbar^{2}/2 \mu R^{2}$ ($\mu$ is the reduced mass). Hereafter, $U_{A0^+_u}(R)$ and $U_{b0^+_u}(R)$ are the locally deperturbed interatomic potentials of $A^{1}\Sigma_{u}^{+}(A0^+_u)$ and $b^{3}\Pi_{0u}^{+}(b0^+_u)$ states, which are represented analytically by the Expanded Morse Oscillator (EMO) functions. The $\xi_{Ab0}(R)$ is the off-diagonal SO coupling function while $A_{01}(R)$ and $A_{12}(R)$ are the $\Omega=1 - 0^{+}$ and $\Omega = 2 - 1$ fine structure splitting functions of the triplet $b^{3}\Pi_u^+$ state, respectively. Both off-diagonal and diagonal SO functions were approximated by the Hulburt-Hirschfelder (HH) potential.

The matrix element $\langle ^1\Sigma^+|H|^3\Pi_1\rangle$ involved in Eq.(\ref{Ham}) is responsible for the indirect 2-nd order (spin-orbit plus spin-rotational) interaction between the $A^{1}\Sigma_{u}^{+}$ and $b^{3}\Pi_{1u}^+$ states through the intermediate $^1\Pi_u$ and $^3\Sigma_u^+$ states. The fitting parameter $\eta_{Ab1}$ is assumed to be independent of $R$.

The refined parameters of the EMO potentials $U^{ab}_{A0^+_u}$, $U^{ab}_{b0^+_u}$, HH spin-orbit functions $\xi^{ab}_{Ab0}$, $A^{ab}_{01}$, $A^{ab}_{12}$, and the $R$-independent parameter $\eta_{Ab1}$ involved in Eq.(\ref{Ham}) were determined iteratively by means of the weighted nonlinear least-square fitting (NLSF) procedure:
\begin{eqnarray}\label{chisquared}
\chi^2=\sum_{j=1}^{N_{Expt}}\left(\frac{E^{Expt}_j-E^{CC}_j}{\sigma^{Expt}_j}\right)^2 + \sum_{j=1}^{N_{ab}}\left (\frac{V^{Emp}_j-V^{ab}_j}{\sigma^{ab}_j}\right)^{2}
\end{eqnarray}
where the rovibronic term values $E^{CC}_j$ and corresponding multi-component vibrational wavefunctions $\mathbf{\Phi}_j(R)$ have been obtained from the iterative solution of the four coupled-channel radial equations:
\begin{eqnarray}\label{CC}
\left(- {\bf I}\frac{\hbar^2 d^2}{2\mu dR^2} + {\bf V}(R;\mu,J) - {\bf I}E^{CC}_j\right)\mathbf{\Phi}_j(R) = 0
\end{eqnarray}
with the conventional boundary $\phi_i(0)=\phi_i(\infty)=0$ and normalization $\sum_{i=1}^{4} P_i=1$ conditions, where $i\in[A^1\Sigma^+_u, b^3\Pi^+_{0u}, b^3\Pi_{1u}, b^3\Pi_{2u}]$.
Here ${\bf I}$ is the identity matrix and $P_i=\langle\phi_i|\phi_i\rangle$ is the fractional partition of the $j$-th level.

The present experimental data set of the $A\sim b$ complex involved in the NLSF procedure (\ref{chisquared}) contains 4503 term values $E^{Expt}_j$ covering the $J$-levels from 4 to 395 and the energy range from 9655 cm$^{-1}$ to 13 630 cm$^{-1}$. The uncertainty of the measured term values $\sigma^{Expt}_j$ could be defined as 0.01 cm$^{-1}$, or slightly less, taking into account the small Doppler effect for the heavy Cs$_2$ molecule.

The 194 term values of the $b^3\Pi^+_{0u}$ state measured in Tsinghua University~\cite{Xie2008} (Tsinghua (LR)) by a monochromator were also included in the present NLSF procedure to extend the experimental region to the bottom of the lower-lying $b$-state. Also 19 term values for $J^{\prime}$ = 1 and 3 from  Ref.~\cite{DanzlPhD2010} were added. These data contained 14 and 5 high resolution term values corresponding to low $[9914, 10~112]$ and high $[12~480, 12~554]$ cm$^{-1}$ energy regions, respectively. The uncertainties $\sigma^{Expt}_j$ of Innsbruck and Tsinghua (LR) data were taken as 0.01 and 1.5 cm$^{-1}$, respectively.

The uncertainties $\sigma^{ab}_j$ of the \emph{ab initio} potentials $U^{ab}_{A0^+_u}$, $U^{ab}_{b0^+_u}$ and of the relevant SO functions $\xi^{ab}_{Ab0}$, $A^{ab}_{01}$, $A^{ab}_{12}$ were estimated by a comparison of the present FSRCC estimates with their previous theoretical counterparts~\cite{Allouche2012, Bai2011}. The initial parameters of the EMO and HH functions required to start the iterative NLSF procedure were borrowed from Ref.~\cite{Bai2011}.

The CC machinery utilized the central five points finite-difference (FD) scheme combined with the analytical mapping procedure~\cite{LeRoy2008} in order to reduce the number of the mesh points required for accurate estimates of eigenvalues belonging to high vibrational levels of the $A\sim b$ complex. The iterative CC calculations were conducted on the interval $R\in [2.8,10.6]$~\AA ~uniformly discretizated by 3000 mesh points of the mapping coordinate. The truncation error of the resulting eigenvalues $E^{CC}$ does not exceed 0.001 cm$^{-1}$ in the energy interval $E_{A\sim b}\in[8000,13~500]$ cm$^{-1}$ of the $A\sim b$ complex. The details on the numerical methods implemented to solve both direct and inverse CC problems can be found elsewhere~\cite{DUO}.

\section{Relativistic electronic structure calculation}\label{zay-relativism}

\subsection{Computational details}
The computational scheme employed for \emph{ab initio} relativistic electronic structure calculation closely resembles that described in details in Ref.~\cite{Zaitsevskii:17}. The basic model was defined by the accurate semilocal shape-consistent two-component pseudopotential of the ``small'' ($1-4s,\,2-4p,\,3-4d$) core of the Cs atom, derived from the valence-shell solutions of the atomic Dirac--Fock--Breit equations with the Fermi nuclear charge model~\cite{Mosyagin:10a}. The contracted Gaussian basis set $[7s\,7p\,5d\,4f\,3g\,1h]$ Cs used to expand the components of one-electron spinors was taken from Ref.~\cite{Zaitsevskii:17}.

The many-electron problem was solved by means of multi-reference Fock space relativistic coupled cluster (FSRCC) method~\cite{Visscher:01, Zaitsevskii:17} using the Fock space scheme ${\rm Cs}_2^{++}\rightarrow {\rm Cs}_2^{+}\rightarrow {\rm Cs}_2$. The cluster operator expansion comprised only single and double excitations (FSRCCSD approximation). The model space defined by 64 Kramers pairs of ``valence'' spinors (lowest virtual solutions of Hartree--Fock-like equations for ${\rm Cs}_2^{++}$) was significantly larger than that used in our previous study~\cite{Zaitsevskii:17}. Numerical instabilities due to the appearance of intruder states were suppressed via introducing adjustable (``dynamic'') shifts of FSRCC energy denominators~\cite{Zaitsevskii:17} in all Fock space sectors, except for the Fermi vacuum one. All calculations were performed using the appropriately modified DIRAC17 program package~\cite{DIRAC:17}.

\subsection{Potential energy curves and spin-orbit matrix elements}

To diminish a systematic $R$-dependent error in the energy calculation~\cite{Zaitsevskii:05,Pazyuk:15} the potential energy curves (PECs) for the excited $(1,2)0^+_u$, $(2)0^-_u$, $(2)1_u$, and $(1)2_u$ states were constructed by adding the FSRCC vertical excitation energies calculated as functions of the internuclear distance, $U_{(n)\Omega^{\pm}_u}(R) - U_{X0^+_g}(R)$, to the highly accurate empirical ground $X$-state potential from Ref.~\cite{Coxon:10}.

Then, resulting relativistic PECs for avoided crossing of $(1,2)0^+_u$ states were converted into the mutually crossing $U^{ab}_{A0^+_u}(R)$, $U^{ab}_{b0^+_u}(R)$ potentials of their ``locally deperturbed'' (SO-decoupled) counterparts $A0^+_u$, $b0^+_u$ and corresponding spin-orbit coupling function $\xi^{ab}_{Ab0}(R)$ through projecting the scalar-relativistic eigenstates $(1)A^1\Sigma^+_u$, $(1)b^3\Pi_u$ on the subspace of strongly coupled $(1,2)0^+_u$ eigenstates of the total relativistic Hamiltonian~\cite{Zaitsevskii:17}. At this stage the many-electron wavefunctions were approximated by their projections onto the FSRCC model space. The resulting matrix elements of the total relativistic electronic Hamiltonian in the basis of projected scalar relativistic states fully incorporate all SO interactions with scalar-relativistic states outside the selected $A^{1}\Sigma_{u}^{+}\sim b^3\Pi_u$ subset. The interatomic potentials $U^{ab}_{A0^+_u}(R)$, $U^{ab}_{b0^+_u}(R)$ and the SO coupling function $\xi^{ab}_{Ab0}(R)$ extracted from the full relativistic calculations should be considered as a complete analog of the locally deperturbed empirical $U^{EMO}_{A0^+_u}(R)$, $U^{EMO}_{b0^+_u}(R)$ potentials and SO coupling $\xi^{emp}_{Ab0}(R)$ function derived in Sec.~\ref{avstol-deperturbation} since both theoretical and empirical functions implicitly absorb the higher order SO interactions with the remote states manifold (including the states embedded into continuum).

The non-equidistant $\Omega$-splitting components $A^{ab}_{01}(R)$, $A^{ab}_{12}(R)$ of the triplet $b^3\Pi_u$ state were determined through the differences $A^{ab}_{01} = U^{ab}_{(2)1_u} - U^{ab}_{b0^+_u}$ and $A^{ab}_{12} = U^{ab}_{(1)2_u} - U^{ab}_{(2)1_u}$, respectively.

\subsection{Transition dipole moments}
\newcommand{\tdm}[0]{M}

Transition electric dipole moments $d_{if}$ between two relativistic adiabatic states ($i$ and $f$) were evaluated using the finite-field scheme~\cite{Zaitsevskii:18}, i.e. the components of
$d_{if}$ were derived from the central finite-difference estimate for the derivative matrix elements in the approximate relation
\begin{equation}\label{fftdm}
(d_{if})_\eta\approx\!\left(E_{f}-E_{i}\right)
\left<\tilde{\Psi}^{\perp\perp}_f (F_\eta)\left|
\frac{\partial \tilde{\Psi}_i(F_\eta)}{\partial F_\eta}\right.\right>
\left|\begin{array}{l}\\_{F=0}\end{array}\right.,
\end{equation}
where $\eta=x,\,y,\,z$, $F$ is the external uniform electric field and $\tilde{\Psi}^{\perp\perp}(F)$ and $\tilde{\Psi}(F)$ denote left and right eigenvectors of the field-dependent
FSRCC effective Hamiltonian acting in the field-independent (constructed assuming $F=0$) model space. Although the calculations involved only the effective Hamiltonian eigenvectors (the model space projections of many-electron wavefunctions), the resulting transition moments implicitly incorporated the bulk of the contributions from the remainder part of these wavefunctions~\cite{Zaitsevskii:98, Zaitsevskii:18}.

\section{Results and Discussion}\label{results}

\subsection{Rovibronic term values and fraction partitions}\label{terms}

\begin{figure}
\includegraphics[scale=1]{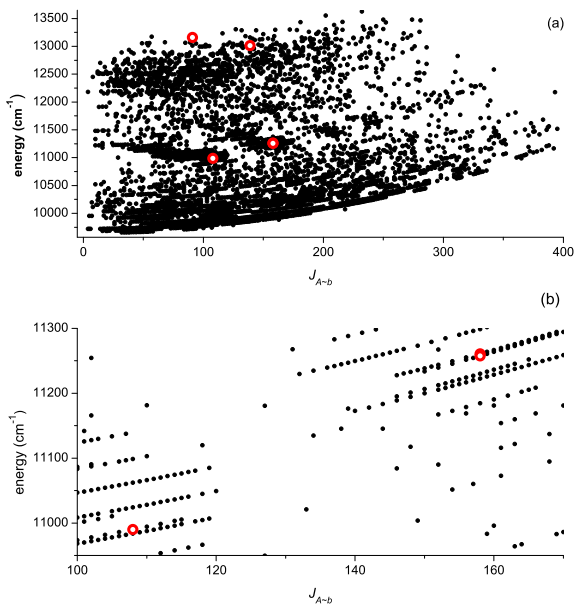}
\caption{(Color online) Data field of the Cs$_2$ $A\sim b$ complex obtained in present work: (a) full scale and (b) zoomed in part. Red open circles denote the levels for which LIF intensity distributions are presented in Section~\ref{intensity}. $J_{A\sim b}\equiv J^{\prime}$ is the rotational quantum number of the $A\sim b$ complex.}\label{Fig_datafield}
\end{figure}

\begin{figure}
\includegraphics[scale=0.4]{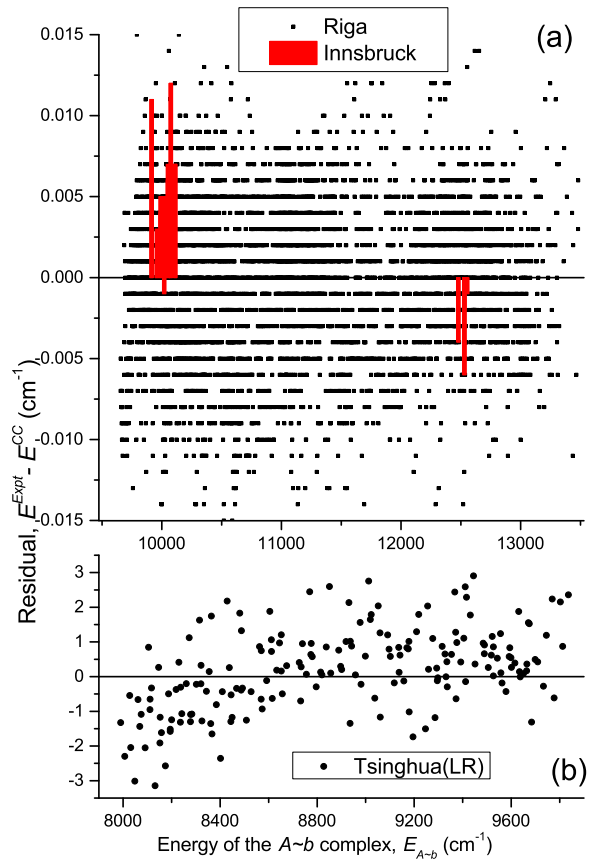}
\caption{(Color online) Residuals (in cm$^{-1}$) between the present least squares fit and experimental term values of Cs$_2$ $A\sim b$ complex obtained in (a) Riga, Innsbruck and (b) Tsinghua (LR).}  \label{Fig_fitresidual}
\end{figure}

Overall data field of presently observed levels of the $A\sim b$ complex contains 4503 term values and is depicted in Fig.~\ref{Fig_datafield}. The quality of the final fit is characterized by the plotted residuals in Fig.~\ref{Fig_fitresidual}, and displayed for the individual data sets in Table~\ref{rmsres}. The current CC model reproduces the Riga and Innsbruck therm values with a standard deviation (SD) of 0.005 cm$^ {-1}$, which is two times less than their estimated experimental uncertainty. The deperturbation model fits the low resolution monochromator Tsinghua (LR) data~\cite{Xie2008} also very well (SD$\sim$ 1.2 cm$^{-1}$) with a small enough mean value (MV) compared with the accuracy of the data. It should be noted that nine outliers of the original Tsinghua (LR) data were excluded from the final fit.

\begin{figure}
\includegraphics[scale=0.4]{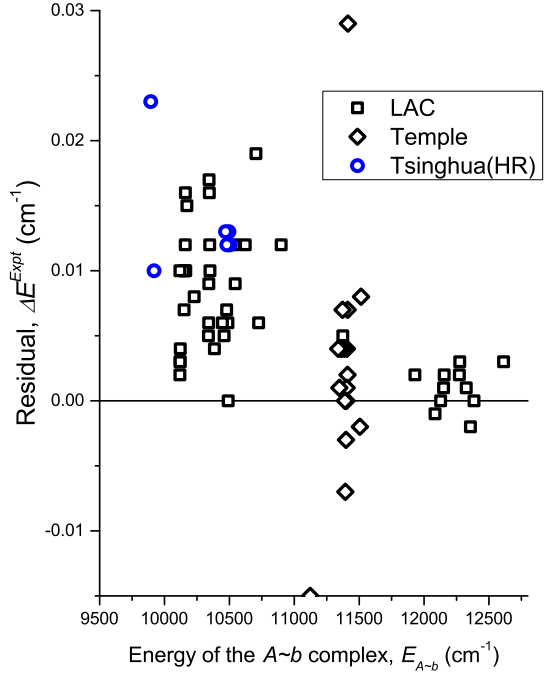}
\caption{(Color online) The difference ($\Delta E^{Expt}$) between the present Riga experimental term values and the ones measured in previous works (LAC~\cite{Verges1987, Amiot2002}, Temple~\cite{Bai2011}, Tsinghua (HR)~\cite{Xie2008}).}\label{Fig_Exptresidual}
\end{figure}

Furthermore, we can see in Table~\ref{rmsres} that the CC model predicts the previously measured in Temple, Tsinghua (HR) and LAC rovibronic term values, which were not included in the fit within their experimental uncertainty while the systematic deviation of about 0.018 cm$^ {-1}$ in Tsinghua (HR) data is clearly observed. The statistic parameters of the present fit correlate well with a direct comparison (see Fig.~\ref{Fig_Exptresidual}) of the Riga term values with previous experimental data available for some rovibrational term values of the $A\sim b$ complex.

\begin{figure}
\includegraphics[scale=0.4]{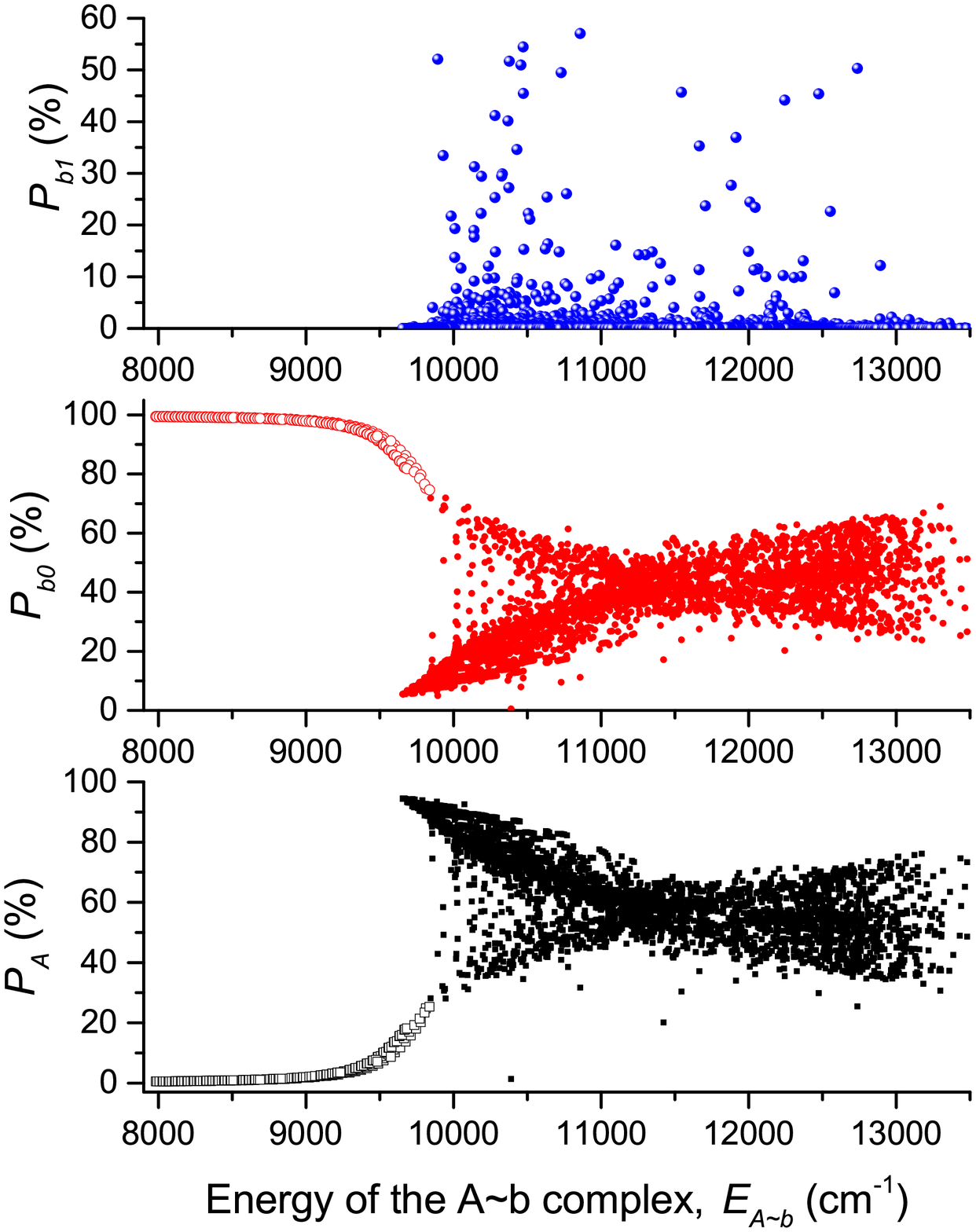}
\caption{(Color online) The fraction partition $P_i=\langle\phi_i|\phi_i\rangle$ ($i\in[A^1\Sigma^+_u, b^3\Pi_{0u}, b^3\Pi_{1u}]$) of the rovibronic levels of the Cs$_2$ $A\sim b$ complex measured in the present work (solid symbols) and previously obtained for low energy range in Ref.~\cite{Xie2008} (open symbols).}\label{Fig_partition}
\end{figure}

The calculated fraction partitions $P_i$ of the rovibronic levels of the $A\sim b$ complex (see Fig.~\ref{Fig_partition}) demonstrate the rapidly growing mixing of the singlet and triplet states above the bottom of the singlet $A$-state. It is interesting, however, that the maximum of the mixing (35-65\%) is observed at the intermediate energy range (about 11 500 cm$^{-1}$), then, the mixing decreases as the excitation energy increases since the overlapping of the vibrational wavefunctions of the interacting states decreases. The present experimental data set contains only about 50 levels having the pronounced $b^{3}\Pi_{1u}$ character with fraction partition $P_{b1u}> 15\%$, and there are no levels with a significant $b^{3}\Pi_{2u}$ character.

\begin{table}
\caption{A comparison of residuals (in cm$^{-1}$) from the present and previous ~\cite{Bai2011} fit of Cs$_{2}$ $A\sim b$ data. Numbers in brackets denote the estimated uncertainty of the measurements in cm$^{-1}$. Data marked by asterisks $^*$ have not been included in the present fit. $N$ is the number of data points ; SD is the standard deviation, and MV is the mean value of the residuals. The data were obtained: Riga - FT LIF spectroscopy (University of Latvia), Tsinghua (LR/HR) - low/high resolution data from Tsinghua University~\cite{Xie2008, Bai2011}, LAC - FT LIF data from Laboratory Aime Cotton~\cite{Verges1987, Amiot2002} and recalculated in Ref.~\cite{Bai2011}. Temple - optical-optical double resonance (OODR) polarization spectroscopy from Temple University~\cite{Bai2011}. Innsbruck - the STIRAP laser assembling of ultracold Cs$_2$ molecules in Innsbruck University~\cite{DanzlPhD2010}.}\label{rmsres}
\begin{tabular}{lrrrl}\hline\hline
     &  $N$ & SD & MV &  \\
\hline
Riga       & 4503 & 0.005 & 0.000 & present\\
(0.01)     & 75   & 0.01  & 0.004 & previous\\
\hline
Innsbruck & 19 & 0.005 & 0.003 & present\\	
(0.01)    & 14 & 0.007 & 0.001 & previous\\
\hline
Tsinghua (LR) & 185 & 1.21 &  0.18 & present\\
(1.50)       & 194 & 1.55 & -0.44 & previous\\
\hline
$^*$LAC   & 338 & 0.005 & 0.005 & present\\
(0.005)   & 340 & 0.015 & 0.007 & previous\\
\hline
$^*$Temple    & 159 & 0.008 & -0.001  & present\\
(0.007)       & 161 & 0.011 & -0.001  & previous\\
\hline
$^*$Tsinghua (HR) & 58 & 0.002 & 0.018 & present\\	
(0.003)          & 58 & 0.009 & 0.003 & previous\\

\hline
\end{tabular}
\end{table}

\subsection{Interatomic potentials and spin-orbit coupling functions}\label{PECSOC}

\begin{figure}
\includegraphics[scale=0.4]{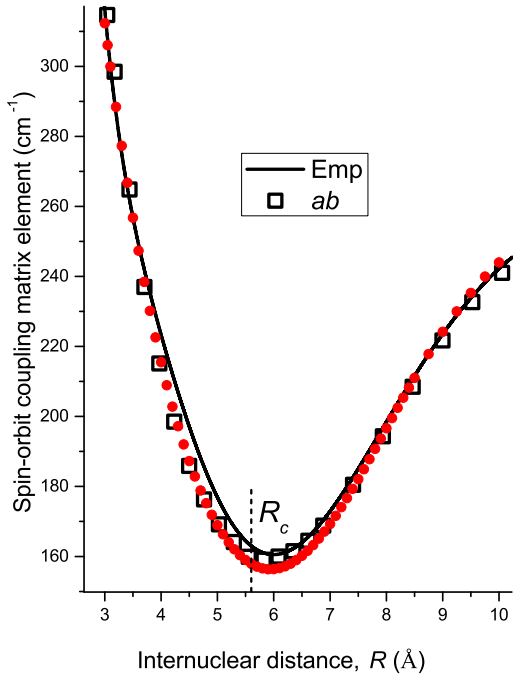}
\caption{(Color online) The comparison of the present empirical and \emph{ab initio} spin-orbit coupling function $\xi_{Ab0}(R)$ between $A^{1}\Sigma_{u}^{+}$ and $b^3\Pi^+_{0u}$ states. The red solid circles denote the previous \emph{ab initio} results obtained as first-order interactions of scalar relativistic states~\cite{Pazyuk2015}. $R_c$ is the crossing point of the locally deperturbed $A0^+_u$ and $b0^+_u$ states.}\label{Fig_SOC}
\end{figure}

\begin{figure}
\includegraphics[scale=0.4]{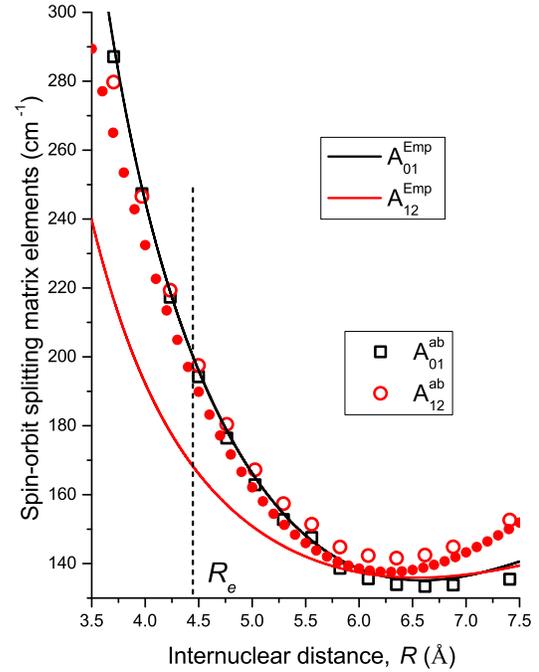}
\caption{(Color online) The comparison of the present empirical and \emph{ab initio} spin-orbit splitting functions $A_{01}(R)$, $A_{12}(R)$ between the $\Omega=0,1,2$ components of the $b^3\Pi_u$ state. The red solid circles denote the equidistant splitting function $A_{01}\equiv A_{12}$ obtained as first-order interactions of scalar relativistic states~\cite{Pazyuk2015}. $R_e$ is the equilibrium distance of the triplet $b$-state.}\label{Fig_SO}
\end{figure}

Table~\ref{EMOP} presents the fitted parameters of the EMO potentials $U_{A0^+_u}(R)$ and $U_{b0^+_u}(R)$ for the locally deperturbed $A0^+_u$ and $b0^+_u$ states belonging to the singlet-triplet $A^{1}\Sigma_{u}^{+}\sim b^{3}\Pi_u$ complex of Cs$_2$. Table~\ref{SOF} presents the resulting parameters of the empirical diagonal $A_{01}(R)$, $A_{12}(R)$ and off-diagonal $\xi_{Ab0}(R)$ spin-orbit functions defined in the HH analytical form. Figures~\ref{Fig_SOC} and ~\ref{Fig_SO} compare the empirical and \emph{ab initio} spin-orbit functions available for the $A\sim b$ complex. The resulting empirical potentials, spin-orbit functions, parameter listings, calculated and observed term values are given in numerical form in the Supplemented Material (SM) ~\cite{EPAPS}. The electronic energies \emph{T$_e$} and equilibrium distances \emph{R$_e$} obtained from the experiment and \emph{ab initio} calculations for the relativistic $(1)0_u^+$, $(2)0_u^\pm$, $(2)1_u$, $(1)2_u$ states and the SO-decoupled $A0^+_u$, $b0^+_u$ states of Cs$_2$ are given in Table~\ref{TeRe}. Overall there is good agreement between the present PECs and their previous counterparts. In particular, Table~\ref{TeRe} demonstrates that though the \emph{ab initio} FSRCC energies are systematically lower than the corresponding empirical curves, however, the differences do not exceed $\sim$ 80-110 cm$^{-1}$. Furthermore, the FSRCC PECs uniformly shifted by 80-90 cm$^{-1}$ diverge from their empirical EMO counterparts only by 30-40 cm$^{-1}$ (approximately one vibrational quanta) in the entire experimental energy range.

The present SO results, see Figs.~\ref{Fig_SOC},~\ref{Fig_SO}, support previous theoretical estimates obtained in the framework of the scalar-relativistic calculations~\cite{Pazyuk2015}. Indeed, the off-diagonal spin-orbit functions $\xi_{Ab0}(R)$ connecting $A^{1}\Sigma_{u}^{+}$ and $b^{3}\Pi_{0u}^+$ states coincide to each other at the crossing point $R_c$ of the interacting states within few wave numbers. The same level of accuracy is achieved for the SO splitting matrix elements $A_{01}(R)$ near the equilibrium distance $R_e$ of the $b^{3}\Pi_u$ state. Moreover, the almost equidistant splitting $A^{ab}_{01}\approx A^{ab}_{12}$ of the $b^{3}\Pi_u$ state is predicted by the present relativistic calculations while the pronounced divergence of the $A^{ab}_{01}(R)$ and $A^{ab}_{12}(R)$ functions at large internuclear distance ($R>8$~\AA~) should be attributed to the increasing impact of the higher-lying $B^1\Pi_u$ and $c^3\Sigma^+_u$ states (see Fig.~\ref{Fig_PEC}). It should be noted that the $A_{12}(R)$ splitting between $\Omega=1$ and $\Omega=2$ components of the $b$-state is empirically not well defined since the experimental term values with a significant $b^{3}\Pi_{2u}$ character were not identified in the present input data.

\begin{table}
\caption{The fitting parameters of Extended Morse Oscillator (EMO) potentials obtained for the locally deperturbed  $A0_u^+$ and $b0^+_u$ states of Cs$_2$:
$U^{EMO} = [T_{dis}-{\mathfrak D}_e] + {\mathfrak D}_e\left [1 - e^{-\beta^{EMO}(R-R_e)}\right ]^2$; $\beta^{EMO}=\sum_{i=0}^{N_a} a_iy^i$, $y=\frac{R^p-R_{\textrm{ref}}^p}{R^p+R_{\textrm{ref}}^p}$. In each case, $N_a$=17, $R_{\textrm{ref}}=5.0$~(\AA) and $p=3$. $R_{e}$ values are in \AA, $T_{dis}$ and ${\mathfrak D}_e$ in cm$^{-1}$, while the polynomial coefficients $a_i$ in 1/\AA. The dissociation limit $T_{dis}(A0^+_u)={\mathfrak D}_e(X)+ E(6^2P_{3/2}) - \xi^{so}_{Cs} =$15197.6606 cm$^{-1}$ and $T_{dis}(b0_u^+)=T_{dis}(A0^+_u) - \xi^{so}_{Cs} =$15012.9812 cm$^{-1}$. ${\mathfrak D}_e(X)=3650.0321$ cm$^{-1}$ is the dissociation energy of the ground $X$-state~\cite{Coxon:10}, $E(6^2P_{3/2;1/2})$ are the fine structure term values of the Cs(6$^2$P) atom~\cite{Atom} and $\xi^{so}_{Cs}=[E(6^2P_{3/2})-E(6^2P_{1/2})]/3$ = 184.6794 cm$^{-1}$ is the spin-orbit constant of the Cs atom.}\label{EMOP}
\begin{tabular}{lcc}
\hline \hline
& $b0_u^+$~state & $A0_u^+$~state \\
\hline
${\mathfrak D}_e$ & 7036.0293 &  5610.9942 \\
$R_e$ & 4.459471 & 5.330011 \\
\hline
$a_{0}$  &  0.522266  &  0.433063 \\
$a_{1}$  &  0.150633  &  0.043114 \\
$a_{2}$  &  0.131508  & -0.016545 \\
$a_{3}$  & -0.222461  &  0.116945 \\
$a_{4}$  & -0.121103  & -0.064999 \\
$a_{5}$  &  2.592014  &  0.105816 \\
$a_{6}$  & -0.554956  &  0.176747 \\
$a_{7}$  & -10.98839  & -0.311183 \\
$a_{8}$  &  7.431177  &  0.012374 \\
$a_{9}$  &  15.57849  &  0.149810 \\
$a_{10}$ & -12.27059  &  0.482162 \\
$a_{11}$ & -0.003271  & -0.871717 \\
$a_{12}$ & -12.01370  &  1.392380 \\
$a_{13}$ & -13.85091  & -0.448932 \\
$a_{14}$ &  22.49664  & -0.099497 \\
$a_{15}$ &  0.072413  & -4.522910 \\
$a_{16}$ &  16.09827  &  0.083157 \\
$a_{17}$ & -10.85234  &  4.296114 \\
\hline
\end{tabular}
\end{table}

\begin{table}
\caption{The fitting parameters for the empirical spin-orbit functions defined in the Hulburt-Hirschfelder (HH) analytical form: $V^{so}_{ij}(R)=\xi^{so}_{Cs}-V^{HH}_{ij}(R)$, where
$V^{HH}_{ij}={\mathfrak D}_e[2e^{-x}-e^{-2x}[1+cx^3(1+bx)]]$ with $x = a(R/R_e - 1)$. ${\mathfrak D}_e$ values are in cm$^{-1}$, $R_e$ are in~\AA; $a$, $b$, $c$ and $\eta_{Ab1}$=0.06384 are dimensionless.}\label{SOF}
\begin{tabular}{lccc}
\hline \hline  
& $\xi^{Emp}_{Ab0}$ & $A^{Emp}_{01}$ & $A^{Emp}_{12}$ \\
\hline
${\mathfrak D}_e$ & 71.1623 & 49.5398 & 48.7475 \\
$R_e$ & 5.99744 & 6.58430 & 6.54553 \\
\hline
$a$   & 2.33165 & 2.72374 & 2.01010 \\
$b$   & 0.57281 & 0.56176 & 0.93778 \\
$c$   & 0.60762 & 0.34739 & 0.40792 \\
\hline
\end{tabular}
\end{table}

\begin{table}
\caption{A comparison of the electronic energies $T_e$ and equilibrium distances $R_e$ available for the relativistic (adiabatic) $(1)0_u^+$, $(2)0_u^\pm$, $(2)1_u$, $(1)2_u$ states and SO-decoupled  $A0^+_u$, $b0^+_u$ states of Cs$_2$. PW - the present work.}\label{TeRe}
\begin{tabular}{clll}
\hline\hline
      & Source & $T_e$(cm$^{-1}$) & $R_e$(\AA)\\
\hline
$A0^+_u$        & Expt.[PW]                  & 9586.64 & 5.330 \\
                & Expt.\cite{Bai2011}        & 9587.12 & 5.329 \\
                & Calc.[PW]                  & 9510    & 5.32  \\
                & Calc.\cite{Zaitsevskii:17} & 9450    & 5.34  \\
                & Calc.\cite{Allouche2012}   & 9601    & 5.22  \\
                & Calc.\cite{Foucrault:92}   & 9710    & 5.24  \\
                & Calc.\cite{Krauss:90}      & 9620    & 5.35  \\
\hline
$b0^+_u$        & Expt.[PW]                  & 7977.18 & 4.459 \\
                & Expt.\cite{Bai2011}        & 7977.85 & 4.458 \\
                & Calc.[PW]                  & 7861    & 4.46  \\
                & Calc.\cite{Zaitsevskii:17} & 7860    & 4.45  \\
\hline
$(2)0_u^+$      & Expt.\cite{Bai2011}        & 9626.64 & 5.290 \\
                & Expt.\cite{Verges1987}     & 9627.06 & 5.292 \\
                & Calc.[PW]                  & 9546    & 5.29  \\
                & Calc.\cite{Zaitsevskii:17} & 9500    & 5.26  \\
                & Calc.\cite{Allouche2012}   & 9667    & 5.17  \\
\hline
$(1)0_u^+$      & Expt.\cite{Bai2011}        & 7960.45 & 4.458 \\
                & Calc.[PW]                  & 7849    & 4.46  \\
                & Calc.\cite{Zaitsevskii:17} & 7850    & 4.46  \\
                & Calc.\cite{Allouche2012}   & 7851    & 4.43  \\
\hline
$(2)0^-_u$      & Expt.\cite{Xie2008}        & 7978.30 & 4.467 \\
                & Calc.[PW]                  & 7867    & 4.46  \\
                & Calc.\cite{Allouche2012}   & 7903    & 4.43  \\
\hline
$(2)1_u$        & Expt.[PW]                  & 8175.20 & 4.481 \\
                & Calc.[PW]                  & 8058    & 4.48  \\
                & Calc.\cite{Allouche2012}   & 8102    & 4.42  \\
                & Calc.\cite{Foucrault:92}   & 8127    & 4.43  \\
                & Calc.\cite{Krauss:90}      & 8470    & 4.51  \\
                & Calc.\cite{Xie2008}        & 8162    & 4.47  \\
\hline
$(1)2_u$        & Expt.[PW]                  & 8341.78 & 4.493 \\
                & Calc.[PW]                  & 8259    & 4.50  \\
                & Calc.\cite{Allouche2012}   & 8321    & 4.43  \\
\hline
\end{tabular}
\end{table}

\subsection{The $\Omega$-doubling effect in the $b^3\Pi^{\pm}_{0u}$ state}\label{doubling}
\begin{figure}
\includegraphics[scale=0.4]{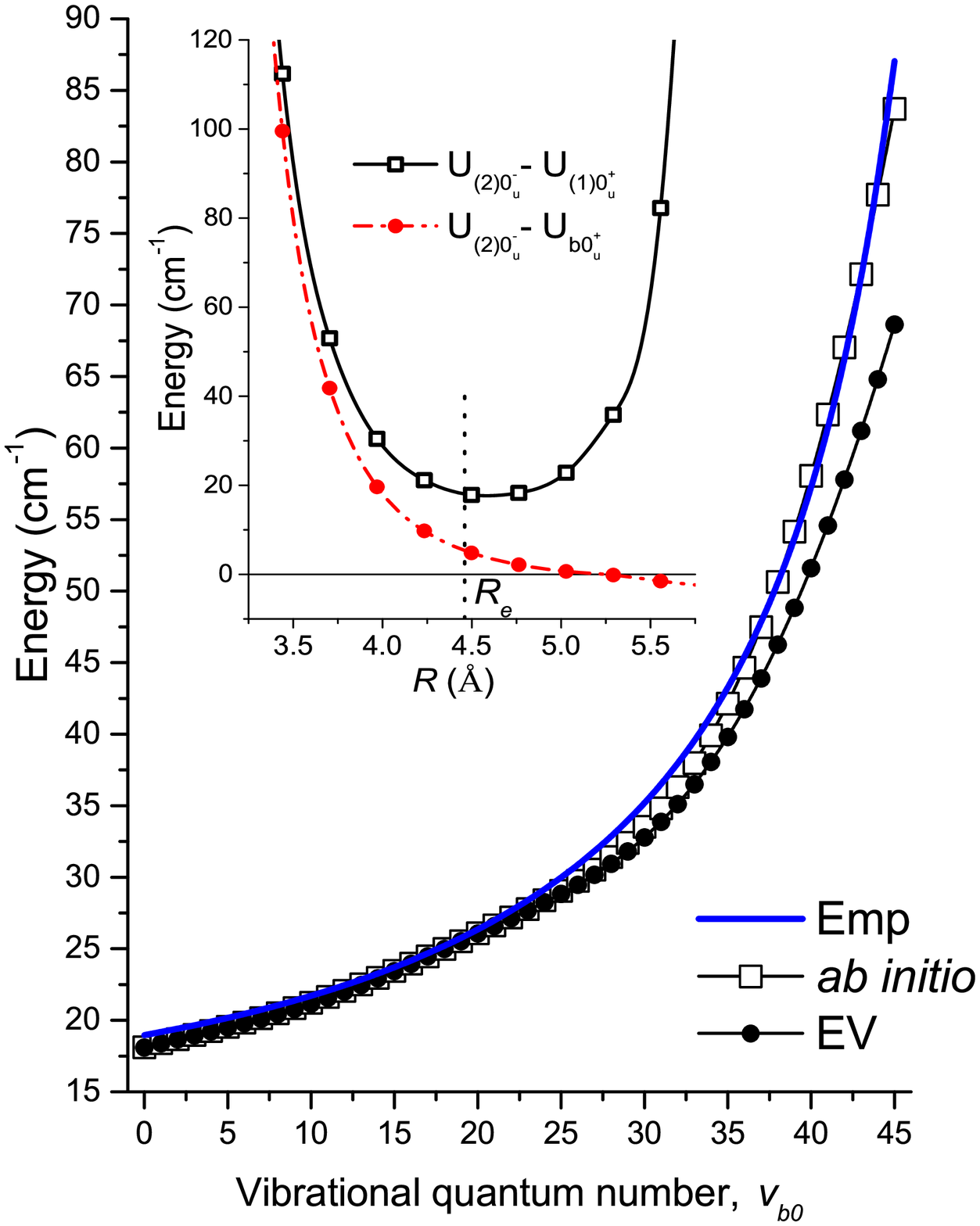}
\caption{(Color online) The comparison of present \emph{ab initio} $\Omega=0^\pm$-splitting energies calculated for the low vibrational levels ($v_{b0}$) of the $b^3\Pi^\pm_{0u}$ state with their empirical  counterparts measured in Ref.~\cite{Xie2008}. The expectation values (EV) were estimated by r.h.s. of Eq. (\ref{splitting}). The inset represents the difference potentials between the relativistic $(2)0^-_u$ and $(1)0^+_u$ states as well as between the relativistic $(2)0^-_u$ and deperturbed $b0^+_u$ state. $R_e$ is the equilibrium distance of the $(2)0^-_u(b0^-_u)$ state.}\label{Fig_splitting}
\end{figure}

To demonstrate the accuracy of the present relativistic PECs the difference of the rovibronic term values $\Delta E^{f/e} = E_{\Omega=0^-_u} - E_{\Omega=0^+_u}$ belonging to the $e$ and $f$ components of the triplet $b^3\Pi^{e/f}_{0u}(b0^{\pm}_u)$ state~\cite{Xie2008} was estimated \emph{ab initio} according to the relation:
\begin{eqnarray}\label{splitting}
E^{vJ}_{(2)0^-_u} - E^{vJ}_{(1)0^+_u} \approx \langle v^J_{(2)0^-_u}|\Delta U^{ab}_{\mp}|v^J_{(2)0^-_u}\rangle,
\end{eqnarray}
where $\Delta U^{ab}_{\mp} = U^{ab}_{(2)0^-_u} - U^{ab}_{(1)0^+_u}$ is the difference of the relativistic $(2)0^-_u$ and $(1)0^+_u$ potentials obtained in the framework of the FSRCC method in Sec.~\ref{zay-relativism}. The required vibrational eigenvalues and eigenfunctions of the relativistic $(1)0^+_u$ and $(2)0^-_u$ states were obtained by solving the single channel radial equation with the corresponding \emph{ab initio} PECs. The resulting $\Delta E^{f/e}$ values depicted in Fig.~\ref{Fig_splitting} agree well with their experimental counterparts measured for low vibrational levels of the  $b^3\Pi^\pm_{0u}$ state~\cite{Xie2008}. The smooth divergence of the expectation values observed for the high vibrational levels should be attributed to a breakdown of the first order perturbation theory used. It should be noted (see the inset in Fig.~\ref{Fig_splitting}) that the interatomic potential $U_{b0_u^+}(R)$ of the deperturbed $b0^+_u$ state is very similar to the adiabatic PEC of the relativistic $(1)0^+_u$ state at $R<R_e$ and becomes very close to the relativistic PEC of the $(2)0^-_u(b0^-_u)$ state at $R>R_e(b0^-_u)$.

\subsection{The $A\sim b\to X$ intensity distributions and rovibronic transition probabilities}\label{intensity}
To validate the reliability of the present CC deperturbation analysis, which has been accomplished so far using only the term values data, relative intensity distributions in the $A\sim b \to X(v^{\prime\prime}_X)$ LIF progressions were measured and compared with their simulated counterparts. This test of the non-adiabatic wavefunctions is especially important in the case of $^{133}$Cs$_2$ where the conventional isotope-substitution analysis is not feasible due to a lack of other stable isotopes.

Relative intensity distributions in LIF progressions were analyzed for a number of upper state levels. Experimental intensities were determined as a peak value of $P$, $R$ lines. The obtained values were corrected according to the spectral sensitivity curve of the InGaAs diode~\cite{hamamatsu}, which shows rather smooth diminishing of the sensitivity almost by a factor of two in the range from 8200 to 9900 cm$^{-1}$. Note that the LIF spectra were measured with a long-pass edge filter FEL1000, which cuts off the LIF at frequencies higher than 10 000 cm$^{-1}$, therefore the transitions to low vibrational levels were not observed.

The corresponding transition probabilities from the rovibronic levels of the $A\sim b$ complex to rovibrational levels of the ground $X$-state were evaluated as
\begin{eqnarray}\label{Iteninterfer}
I_{A\sim b\to X}^{calc} &\sim& \nu^4|\langle \phi_A^{J^{\prime}}|d_{AX}|v_X^{J^{\prime\prime}}\rangle|^2,\\
\nu &=& E^{CC}_{A\sim b}(J^{\prime})-E_{v_X}(J^{\prime\prime}=J^{\prime}\pm 1),\nonumber
\end{eqnarray}
where the rovibrational eigenvalues $E^{CC}_{A\sim b}(J^{\prime})$ and the singlet $A$-component $|\phi_A^{J^{\prime}}\rangle$ of non-adiabatic eigenfunctions for the complex were obtained from the solution of the CC equations (\ref{CC}) with the present empirical potentials and spin-orbit functions. The interatomic potential needed to calculate the adiabatic energies $E_{v_X}(J^{\prime\prime})$ and wavefunctions $|v_X^{J^{\prime\prime}}\rangle$ of the ground $X$-state was borrowed from Ref.~\cite{Coxon:10}.

\begin{figure}
\includegraphics[scale=0.4]{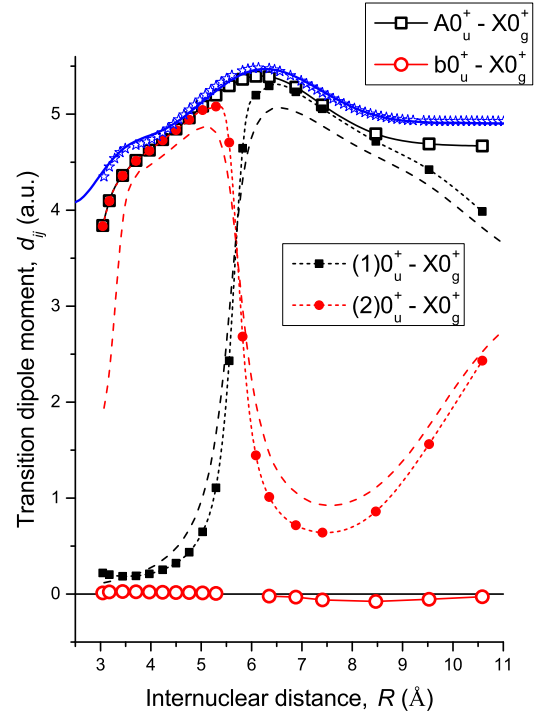}
\caption{(Color online) The \emph{ab initio} relativistic $(1,2)0^+_u-X0^+_g$ electronic transition dipole moments obtained in the framework of the present finite-field FSRCC method (solid symbols) and in Ref.~\cite{Allouche2012} (dashed lines), respectively. Their SO-decoupled $A0^+_u-X0^+_g$ and $b0^+_u-X0^+_g$ counterparts (open squares) were estimated by means of the unitary transformation (\ref{tranunitary}). The spin-allowed $A^1\Sigma^+_u - X^1\Sigma^+_g$ transition dipole moments were obtained in Refs.\cite{Allouche2012, Pazyuk2016} (open stars and solid line, respectively) using the scalar relativistic electronic wave functions.}\label{Fig_ETDM}
\end{figure}

The $d_{AX}(R)$ function involved in Eq. (\ref{Iteninterfer}) is the SO-decoupled $A0^+_u-X0^+_g$ transition dipole moment (see Fig.~\ref{Fig_ETDM}) which was evaluated by means of the unitary transformation of the relativistic $d_{(1,2)0^+_u-X0^+_g}$ moments obtained in Sec.~\ref{zay-relativism} in the framework of the finite-field FSRCC method (see Fig.~\ref{Fig_ETDM}):
\begin{eqnarray}\label{tranunitary}
d_{AX} &=&  \cos \theta d_{1X} - \sin \theta d_{2X} \\
d_{bX} &=&  \sin \theta d_{1X} + \cos \theta d_{2X},\nonumber
\end{eqnarray}
where the transformation angle~\cite{Field} $\theta(R)$ is the function of the \emph{ab initio} SO coupling matrix element $\xi^{ab}_{Ab0}(R)$ and the corresponding SO-decoupled potentials $U^{ab}_{A0^+_u}(R)$, $U^{ab}_{b0^+_u}(R)$:
\begin{eqnarray}\label{transform}
\theta = \frac{1}{2}\arctan\frac{2\xi^{ab}_{Ab0}}{U^{ab}_{A0^+_u}-U^{ab}_{b0^+_u}}.
\end{eqnarray}

The resulting $d_{AX}$ function corresponding to the spin-allowed $A^{1}\Sigma_{u}^{+}-X^{1}\Sigma_{g}^{+}$ transition is found to be very close to its scalar-relativistic counterparts obtained in Refs.~\cite{Allouche2012, Pazyuk2016} using the ECP-CPP-CI method. It should be noted that the ECP-CPP-CI model~\cite{Allouche2012} based on the full configuration interaction (CI) treatment of two-valence-electron problem defined by the large-core two-component relativistic pseudopotentials (ECP) of Cs atoms and the core-valence correlation treatment through semiempirical core-polarization potential (CPP) provides a good approximation to the transition dipole moment functions between the relativistic (adiabatic) states. The systematic divergence of the ECP-CPP-CI $A^1\Sigma_u^+-X^1\Sigma_g^+$ and transformed FF FSRCC $A0_{u}^{+}-X0_{g}^{+}$ transition moments, observed at small and large internuclear distance, is comparable to and even less than the typical uncertainty of the measured intensities.

\begin{figure}
\includegraphics[scale=0.4]{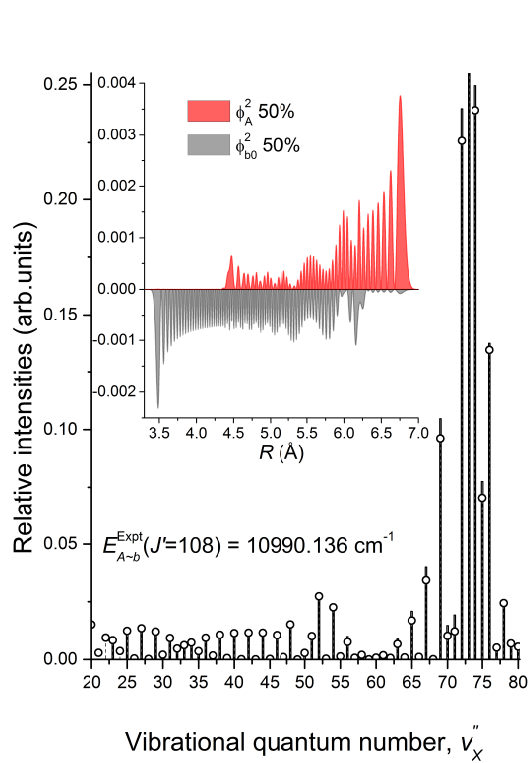}
\caption{(Color online) The experimental (vertical bars) and calculated (empty circles) relative intensity distributions in the vibrational $A\sim b \to X(v^{\prime\prime}_X)$ LIF progression originating from the fully mixed (50\%/50\%) level of the $A\sim b$ complex. Maximal line intensity at $v^{\prime\prime}_X$ = 73 is normalized to one and cut at intensity 0.25 for visibility.  The inset represents the calculated density distribution of the corresponding multi-component vibrational wavefunctions of the singlet $A^1\Sigma^+_u$ (upper part) and triplet $b^3\Pi^+_{0u}$ (lower part) states as dependent on internuclear distance.}\label{Fig_Int50_50}
\end{figure}

\begin{figure}
\includegraphics[scale=0.49]{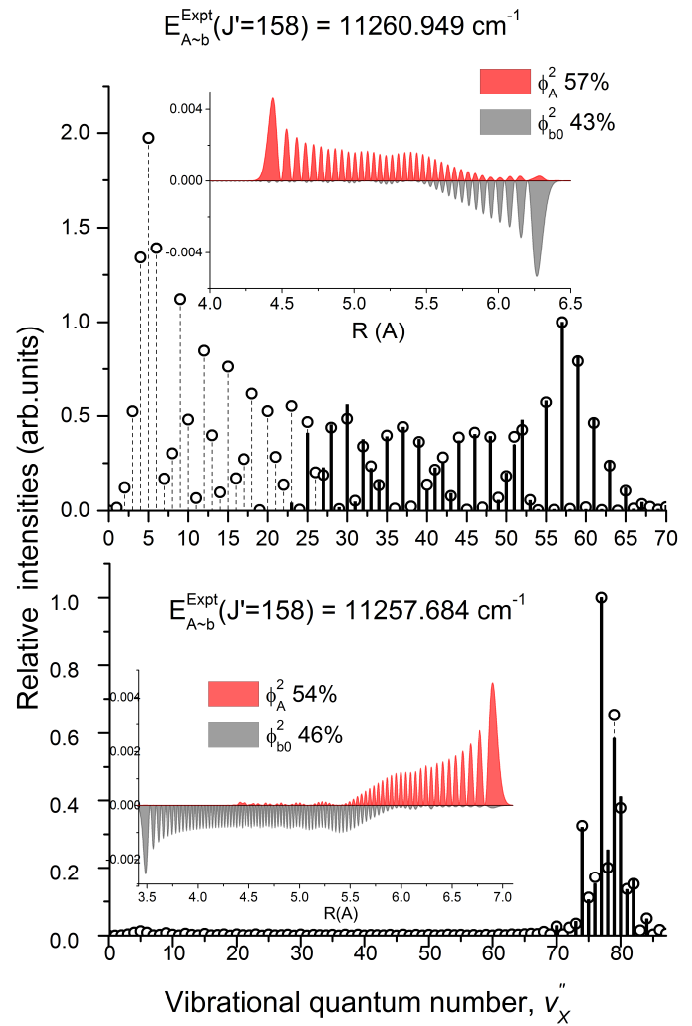}
\caption{(Color online) The experimental (bold bars) and calculated (empty circles) relative intensity distributions in the $A\sim b \to X(v^{\prime\prime}_X)$ LIF (upper panel)  and CIF (lower panel) progressions starting from the close-lying rovibrational levels (with $J^\prime$=158) of the $A\sim b$ complex. The optical filter cuts off LIF below  $v^{\prime\prime}_X$ = 25. The insets represent the density distribution of the multi-component wavefunctions of the $A^1\Sigma^+_u$ (upper part) and $b^3\Pi^+_{0u}$ (lower part) states.}\label{Fig_IntCIF}
\end{figure}

\begin{figure}
\includegraphics[scale=0.4]{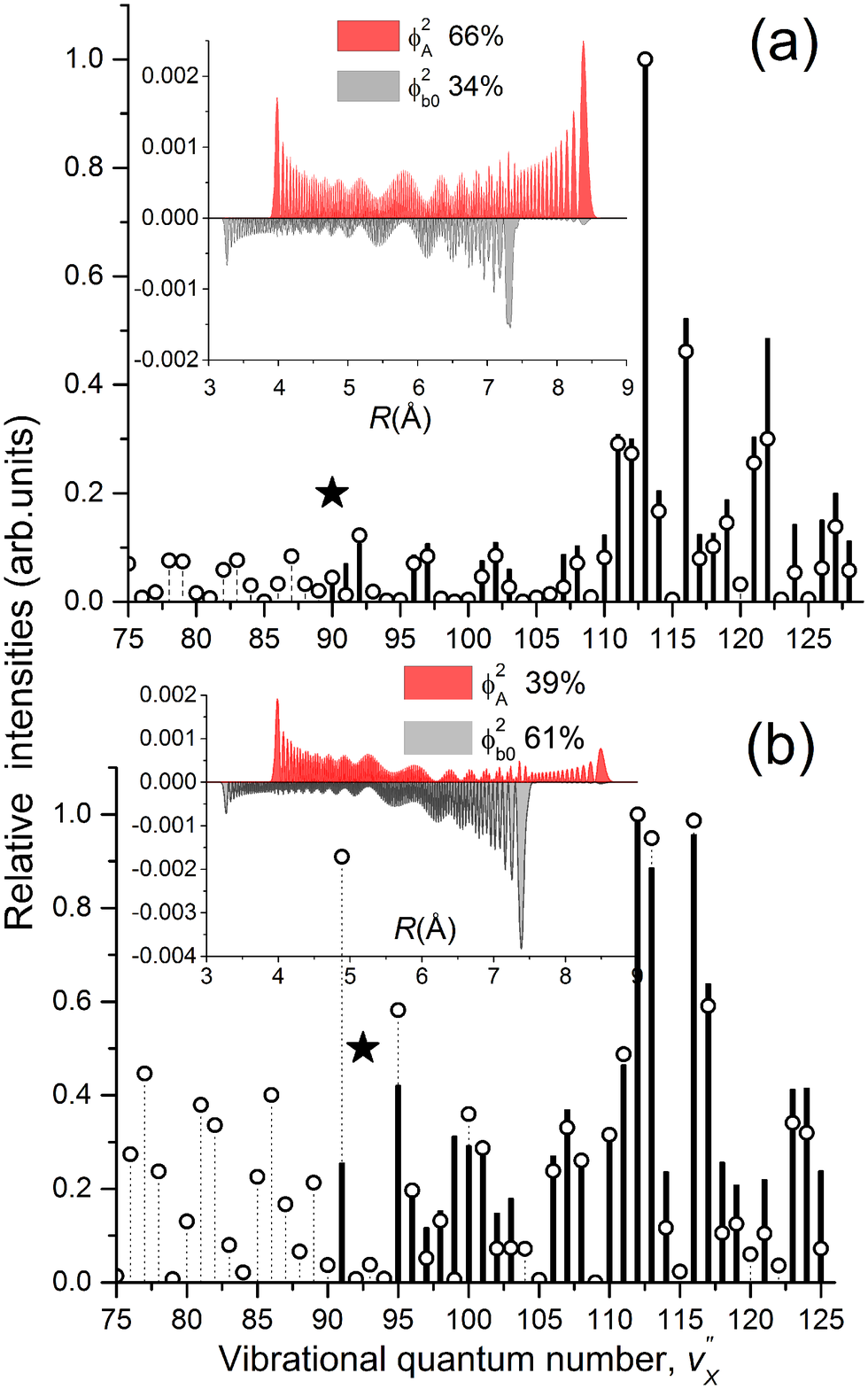}
\caption{(Color online) A comparison of the experimental (bold bars) and calculated (empty circles) relative intensity distributions in the $A\sim b \to X(v^{\prime\prime}_X)$ LIF progressions originating from the high vibrational levels of the $A\sim b$ complex: (a) $J^\prime$=139, $E_{A\sim b}=$13 012.260 cm$^{-1}$, (b) $J^\prime$=91, $E_{A\sim b}=$13 162.000 cm$^{-1}$. The optical filter cuts off LIF below  $v^{\prime\prime}_X$ = 90, see solid stars. The insets represent the density distribution of the non-adiabatic vibrational wavefunctions of the $A^1\Sigma^+_u$ (upper part) and $b^3\Pi^+_{0u}$ (lower part) states.}\label{Fig_Inthighv}
\end{figure}

For a comparison of experimental and calculated relative intensity distributions, the averaged values of $P$ and $R$ line intensities were used. Figures ~\ref{Fig_Int50_50},~\ref{Fig_IntCIF} and ~\ref{Fig_Inthighv} represent some examples for the cases when a strong mixing in upper state levels takes place (see marked levels in Fig.~\ref{Fig_datafield}). These examples show overall excellent agreement between the measured and calculated intensity distributions in LIF progressions. The calculated density distribution of the corresponding multi-component vibrational wavefunctions of the $A\sim b$ complex clearly demonstrates the dramatic changes in the nodal structure of the non-adiabatic vibrational wavefunctions in comparison with their conventional adiabatic (single-channel) counterparts~\cite{PCCP:10}. In particular, as is seen from the inset of Fig.~\ref{Fig_Int50_50}, the density of the singlet component of non-adiabatic wavefunction $|\phi_A|^2$ belonging to the strongly mixed level of the $A\sim b$ complex is mainly distributed near the right turning point and, hence, the corresponding band intensities of LIF are localized in the squeezed region of the vibrational levels $v^{\prime\prime}_X=73\pm 1$ of the ground state. For LIF (upper panel) and CIF (lower panel) progressions observed from the close-lying rovibrational levels (with $J^\prime$=158) of the $A\sim b$ complex the alternative picture takes place (see Fig.~\ref{Fig_IntCIF}). In the LIF case, the density function $|\phi_A|^2$ is distributed from the left turning point to the middle range of $R$, while in the CIF case from the middle $R$ to the right turning point. Thus, the broad LIF progression ends at $v^{\prime\prime}_X$ around 65 whereas the narrow CIF progression just starts from $v^{\prime\prime}_X> 70$ and possesses a sharp maximum at $v^{\prime\prime}_X$ around 77. For very high vibrational levels of the $A\sim b$ complex ($E_{A\sim b}>$13000 cm$^{-1}$) the mutual perturbation decreases and singlet component of wavefunctions starts to be distributed in entire classical range from the left to the right turning points (see Fig.~\ref{Fig_Inthighv}), therefore the corresponding long $A\sim b \to X(v^{\prime\prime}_X)$ LIF progressions extend to very high $v^{\prime\prime}_X$ of the ground state.

\section{Concluding remarks}\label{remarks}
Fourier-transform LIF spectra of $A\sim b\rightarrow X$ transitions were recorded for the $^{133}$Cs$_2$ molecule. Overall 4503 rovibronic term values covering the energy range $E_{A\sim b}\in [9655, 13630]$ cm$^{-1}$ and rotational levels $J_{A\sim b}\in [4, 395]$ of the strongly coupled $A^1\Sigma^+_u$ and $b^3\Pi_u$ states were determined with an uncertainty of 0.01 cm$^{-1}$. These data were combined with experimental data from other sources and were simultaneously involved in the direct deperturbation analysis performed in the framework of the inverted coupled-channels approach. The deperturbed potential energy curves of the interacting $A^1\Sigma^+_u$ and $b^3\Pi^+_{0u}$ states and the relevant spin-orbit coupling functions reproduce the FT spectroscopy rovibronic term values with a standard deviation of 0.005 cm$^{-1}$. The excellent agreement between the experimental relative intensity distributions measured in the $A\sim b\rightarrow X(v^{\prime\prime}_X)$ LIF progressions and their theoretical counterparts unambiguously supports the non-adiabatic eigenfunctions of the $A\sim b$ complex and relativistic structure calculations, including transition dipole moments, of Cs$_2$.

\begin{acknowledgments}
Moscow team acknowledges the support from the Russian Government Base Funding No AAAA-A16-116052010077-8: "Quantum chemistry modeling and laser-induced breakdown spectrometry".
Riga team acknowledges the support from the Latvian Council of Science project "Determination of structural and dynamic properties of alkali diatomic molecules for quantum technology applications", project No. LZP2018/1-0020 and from the University of Latvia Base Funding No A5-AZ27; A.K. acknowledges the support from the Post-doctoral Grant No 1.1.1.2/16/I/001, proposal No 1.1.1.2/I/16/068.
\end{acknowledgments}.

\end{document}